\documentclass[twoside,11pt]{article}
\usepackage{cite}                         
\usepackage{citesort}                     
\usepackage[dvips]{color}               
\usepackage{epsf}
\usepackage{graphicx}
\usepackage{amsmath}

\def\NPA{{\em Nucl. Phys.} A}

\def\NPB{{\em Nucl. Phys.} B}

\def\PLB{{\em Phys. Lett.} B}

\def\PRD{{\em Phys. Rev.} D}
\def\PRC{{\em Phys. Rev.} C}

\def\beq{\begin{equation}}
\def\eeq{\end{equation}}
\def\bea{\begin{eqnarray}}
\def\eea{\end{eqnarray}}
\def\beqa{\begin{equation}\begin{array}{l}}
\def\eeqa{\end{array}\end{equation}}
\def\eqlab#1{\label{eq:#1}}
\def\figlab#1{\label{fig:#1}}

\def\Eqref#1{Eq.~(\ref{eq:#1})}

\def\Figref#1{Fig.~\ref{fig:#1}}

\def\half{\mbox{\small{$\frac{1}{2}$}}}

\def\quarter{\mbox{\small{$\frac{1}{4}$}}}

\def\barr{\left(\begin{array}{c}}
\def\earr{\end{array}\right)}
\def\bmat{\left(\begin{array}{cc}}
\def\emat{\end{array}\right)}
\def\al{\alpha}
\def\be{\beta}
\def\ga{\gamma} 
\def\de{\delta} \def\De{\Delta}
\def\veps{\varepsilon}  \def\eps{\epsilon}

\def\la{\lambda}

\def\si{\sigma} 
  
\def\w{\omega}

\def\pa{\partial}

\def\pa{\partial}

\def\nn{\nonumber}

\def\lag{{\mathcal L}}

\def\mathscr{\mathcal}

\def\arctg{{\rm arctg}}

\def\3d{3-D}

\def\ol#1{\overline{#1}}
\def\amm{a.m.m.}

\begin{document}

\title{\vspace*{-0.5cm}\hfill {\large WM-04-123}\\[-4mm]
\hfill {\large JLAB-05-04-306}\\
\vspace*{1.5cm} Some electromagnetic properties of the nucleon from \\
Relativistic Chiral Effective Field Theory\footnote{\small Invited seminar at
the 26th Course of the International Erice School of Nuclear Physics: {\it Lepton
                    Scattering and the Structure of Hadrons and Nuclei}, Erice,
                  Italy, 16--24 Sep 2004. To appear in Prog.~Nucl.~Part.~Phys.~54.
}}
              
\author{
{ Vladimir Pascalutsa}\\ 
{\em Physics Department, The College of William \& Mary, Williamsburg, VA
23187, USA} \\
{\em Theory Group, Jefferson Lab, 12000 Jefferson Ave, Newport News, VA 23606, USA} } 

\date{(December 1, 2004)}
\maketitle

\begin{abstract} \noindent
Considering the magnetic moment and polarizabilities
of the nucleon we emphasize the need for {\it relativistic} 
chiral EFT calculations. Our relativistic calculations are
done via the forward-Compton-scattering sum rules, thus ensuring the
correct analytic properties. The results obtained in this way
are equivalent to the usual loop calculations, provided no
heavy-baryon expansion or any other manipulations
which lead to a different analytic structure (e.g., infrared
regularization) are made.   
The Baldin sum rule can directly be applied to 
calculate  the sum of nucleon polarizabilities. In contrast,
the GDH sum rule is practically unsuitable for calculating 
the magnetic moments. The breakthrough is
achieved by taking the derivatives of the sum rule with respect to the anomalous magnetic
moment.
As an example, we apply the derivative of the GDH sum rule to
the calculation of the magnetic moment in QED and reproduce the famous Schwinger's correction
from a tree-level cross-section calculation.  
As far as the nucleon properties are concerned, we focus on two issues:
1) chiral behavior of the nucleon magnetic moment and 2) reconciliation of the
chiral loop and $\Delta$-resonance contributions to the nucleon magnetic polarizability.
\end{abstract}
\thispagestyle{empty}

\newpage
\section{GDH sum rule and its derivatives}
Consider the elastic scattering of a photon on a target with spin $s$ (real Compton scattering).
The forward-scattering amplitude of this process is characterized 
by $2s+1$ scalar functions which depend on a single kinematic
variable, e.g., the photon energy $\w$. In the low-energy limit
each of these functions corresponds to an electromagnetic moment ---
charge, magnetic dipole, electric quadrupole, etc. --- of the target.
For example in the case of the nucleon, the 
forward Compton amplitude is generally written as
\begin{eqnarray}
\label{DDeq2.2.2}
T(\w) = \vec \veps'\cdot\vec \veps\,f(\w)+
i\,\vec \sigma\cdot(\vec\veps'\times\vec\veps)\,g(\w)\, ,
\end{eqnarray}
where $\vec \veps$, $\vec \veps'$ is the polarization vector of the incident
and scattered photon, respectively; $\vec \sigma$ are the Pauli matrices
representing the dependence on the nucleon spin. The two scalar functions
have the following low-energy expansion,
\begin{subequations}
\eqlab{let}
\begin{eqnarray}
f(\w) & = & -\frac{e^2}{4\pi M} + (\alpha_E+\beta_M)
\,\w^2+ {\mathcal{O}}(\w^4) \ , \label{DDeq2.2.12} \\
g(\w) & = & -\frac{e^2\kappa^2}{8\pi M^2}\,\w +
\gamma_{0}\w^3 + {\mathcal{O}}(\w^5) \ , \label{DDeq2.2.13}
\end{eqnarray}
\end{subequations}
hence in the low-energy limit they are given in terms of the nucleon charge $e$ 
and the anomalous magnetic moment (a.m.m.) $\kappa$. The next-to-leading order
terms are specified by the nucleon electric ($\al_E$), magnetic ($\be_M$), and
forward spin ($\ga_0$) polarizabilities.

To derive the sum rules (SRs) for these quantities one assumes  the scattering amplitude
is an {\it analytic} function of $\w$ everywhere but the real axis\footnote{Resonance poles
may occur but lie on the second Riemann sheet.}. This allows us to write the real parts of
functions  $f(\w)$ and $g(\w)$ as a {\it dispersion
integral} of their imaginary parts. The latter, on the other hand, can be related to the
total photoabsorption cross-sections by using the {\it optical theorem},
\begin{subequations}
\begin{eqnarray}
\mbox{Im}\ f(\w) & = & \frac{\w}{8\pi}
\left[\sigma_{1/2}(\w)+\sigma_{3/2}(\w)\right] \,, \\
\mbox{Im}\ g(\w) & = & \frac{\w}{8\pi}
\left[\sigma_{1/2}(\w)-\sigma_{3/2}(\w)\right] \,,
\label{DDeq2.2.6}
\end{eqnarray}
\end{subequations}
where $\sigma_{\la}$ is the double-polarized total cross-section of the
photoabsorption processes. Averaging over the polarization of initial
particles gives the total unpolarized cross-section,
 $\sigma_T=\half ( \sigma_{1/2}+\sigma_{3/2})$.

Finally one uses the {\it crossing symmetry}, meaning that the
Compton amplitude  of
Eq.~(\ref{DDeq2.2.2}) must be invariant under 
$\varepsilon'\leftrightarrow\varepsilon$,
$\w\rightarrow-\w$, and hence $f$ is an even and
$g$ an odd function of energy:$f(\w) = f(-\w)$, $g(\w) =-g(-\w)$.

Going through these steps one arrives at the following result (see, {\it e.g.}, \cite{DPM02}
for more details):
\bea
\label{DDeq2.2.8}
f(\w) & = & \frac{1}{2\pi^2}\,
\int_{0}^{\infty}\frac{\sigma_T(\w')}
{\w'^2-\w^2 - i\eps}\,\w'^2\, d\w'\ . \\
\label{DDeq2.2.9}
 g(\w) &=& -\frac{\w}{4\pi^2}\,
\int_{0}^{\infty}\frac{\De\sigma(\w')}
{\w'^2-\w^2 - i\eps}\,\w'\,d\w'\ ,
\eea
with $\De\sigma\equiv \sigma_{3/2}-\sigma_{1/2}$.
These relations can be expanded in energy to obtain the SRs for the
different static properties introduced in \Eqref{let}.
In this way we can obtain the Baldin SR:
\begin{eqnarray}
\label{DDeq2.2.14}
\alpha_E + \beta_M = \frac{1}{2\pi^2}\,
\int_{0}^{\infty}\,\frac{\sigma_T(\w')}{\w'^2}
\,d\w'\ ,
\end{eqnarray}
the Gerasimov-Drell-Hearn (GDH)
SR:
\begin{eqnarray}
\label{DDeq2.2.15}
\frac{e^2\kappa^2}{2M^2}=\frac{1}{\pi}
\int_{0}^{\infty}\,\frac{\De\sigma(\w)}{\w}\,d\w \, ,
\end{eqnarray}
and a SR for the forward spin
polarizability:
\begin{eqnarray}
\label{DDeq2.2.16}
\gamma_0= \,-\,\frac{1}{4\pi^2}\,\int_{0}^{\infty}\,
\frac{\De\sigma(\w)}
{\w^3}\,d\w\ .
\end{eqnarray}

As we all now know, impressive experimental programs to measure 
the total photoabsorption cross-sections of the nucleon have recently been carried out
at ELSA and MAMI (for a review see Ref.~\cite{Grab}). 
These measurements are needed for an empirical test of the GDH SR,
as well as for phenomenological estimates of $\al_E+\be_M$ and
$\gamma_0$ via the other two SRs. The GDH SR is particularly interesting because
both the left- and right-hand-side of this SR can reliably be measured,
thus providing a test of the fundamental principles (such as unitarity
and analyticity) which go into its derivation. 

Testing {\it theories} by using these SRs could also be fun and even useful
in instances when the consistency of the theory is not transparent. Let us, for example, have a look
at the left- and right-hand-sides of the GDH SR for the electron in QED. To lowest
order in the fine-structure constant, $\al=e^2/4\pi$, the photoabsorption cross-section is given 
by the tree-level Compton scattering cross-section~\cite{wk}:
\beq
 \De\sigma(\omega) = \frac{2\pi\alpha^2}{M\w}\left[2+{2 \w^2\over
(M+2\w)^2} - \left(1+{M\over
\w}\right)\,\ln\left(1+\frac{2\w}{M}\right)  \right] + O(\al^3) \label{eq:gdh} .
\eeq
On the other hand, the one-loop contribution to the electron \amm\ is
of order $\al$ and therefore the {\it lhs} has no contribution of $\al^2$. Fortunately,
the GDH integral over the tree-level cross-section \Eqref{gdh} {\it vanishes}, and thus, 
at this order, everything works out:
\beq
 0 = 0
\eeq
as one could expect for such a fortunate theory as QED.
At order $\al^3$ the {\it lhs} receives 
the contribution in the form of  Schwinger's correction: $\kappa=\al/2\pi$.
The calculation of  cross sections at this order is quite a formidable task as it requires
the knowledge of Compton scattering amplitude to one-loop, 
inclusion of the pair-production channel, and so on,
cf.~\cite{DiV01}. Instead, I want to follow a much simpler way to do calculations at this 
level~\cite{PHV04}. 

Let us introduce a `classical' (or 'trial') value of the \amm, $\kappa_0$. At the level of the Lagrangian
this amounts to introducing a Pauli term for our spin-1/2 field: 
\beq
{\cal L}_{\mbox{\small Pauli}}=
(i\kappa_0/4M) \, \bar \psi\,  \si_{\mu\nu}\, \psi\, F^{\mu\nu}\,,
\eeq
 where $F$ is the electromagnetic
field tensor and $\si_{\mu\nu}=(i/2) [\ga_\mu,\ga_\nu]$. 
In the end we can put $\kappa_0$ equal to zero, but for now the total value
of the \amm\ is $\kappa=\kappa_0 +\delta \kappa$, with $\de\kappa$ being 
the quantum effects. Note that $\de\kappa$ and the total cross-sections become
explicitly dependent on $\kappa_0$. To get something new out of this we need to start
taking derivatives of the GDH SR with respect to $\kappa_0$:
\bea
(4\pi^2\alpha/M^2)\, \kappa \,\kappa' &=&
\int_{0}^\infty \!  \De\si'(\w)\, \frac{d\w}{\w}\,,\\
(4\pi^2\alpha/M^2)\, ( \kappa'^2+\kappa\,\kappa'')  &=&
\int_{0}^\infty \!  \De\si''(\w)\,\frac{d\w}{\w}\,,
\eea
and so on. Now observe that to lowest order in $\al$ these relations simply read as
\beq
(4\pi^2\alpha/M^2)\, n\,\kappa^{(n-1)} = \int_{0}^\infty \!\!  \De\si^{(n)}(\w)\, \frac{d\w}{\w}\,
\eeq
where $n$ is the order of the derivative with respect to $\kappa_0$. This allows us in principle 
to compute $\kappa$ to order $\al^k$ by using $k$th, $(k-1)$th, etc., derivatives of
the cross-section computed, respectively, to order $\al^{k+1}$,  $\al^{k}$, etc. 
In this way, to lowest order we have the following sum rule:
\beq
\eqlab{linsr}
(4\pi^2\alpha/M^2)\, \kappa  =
\int_{0}^\infty \!  \left. \De\si'(\w)\right|_{\kappa_0=0}\, \frac{d\w}{\w}\,.
\eeq
The striking feature of this sum rule is the {\it linear}
relation between the \amm\ and the photoabsorption cross section,
in contrast to the GDH SR where the relation is quadratic.  This restores the 
``balance of difficulty'' in the two methods of calculating this quantity:
the sum rule or the usual loop technique.

Although 
the cross-section quantity $\De\si'$ is not an
observable, it is very
clear how to determine it within a given theory.
The first derivative of the tree-level cross-section with respect to $\kappa_0$, at $\kappa_0=0$, 
in QED takes the form~\cite{PHV04}: 
\beq
\left. \De\si'(\w)\right|_{\kappa_0=0} = \frac{2\pi\alpha^2}{M\w }\left[ 6-{2M\w
\over (M+2\w)^2} -\left(2+{3M\over \w}\right)\,\ln\left(1+\frac{2\w}{M}\right) \right] .
\label{eq:gdh2} 
\eeq
It is then not difficult to find that
\begin{equation}
\frac{1}{\pi}\int\limits_0^\infty \!\left. \De\si'(\w)\right|_{\kappa_0=0} \,{d\omega\over \omega }
=
\frac{2\al^2}{M^2} \,.
\end{equation}
Substituting this result in the linearized SR, \Eqref{linsr}, we obtain $\kappa=\al/2\pi$.
Thus, the Schwinger's one-loop result is reproduced here by computing
only a (derivative of the) tree-level Compton scattering cross-section and then performing a GDH integral.


%

\section{Magnetic moments and their chiral extrapolation}
Consider now the theory of nucleons
interacting with pions via pseudovector coupling: \beq {\cal
L}_{\pi NN} = \frac{g}{2 M}\, \bar\psi \,\ga^\mu \,\ga^5 \,\tau^a
\,\psi \,\pa_\mu \pi^a, \eeq where $g$ is the pion-nucleon
coupling constant, $\tau^a$ are isospin Pauli matrices, $\pi^a$ is
the isovector pion field. For our purposes this Lagrangian is sufficient
to obtain the leading order results of chiral perturbation theory.
 \begin{figure}[b,h,t,p]
\centerline{
  \epsfxsize=12cm
  \epsffile{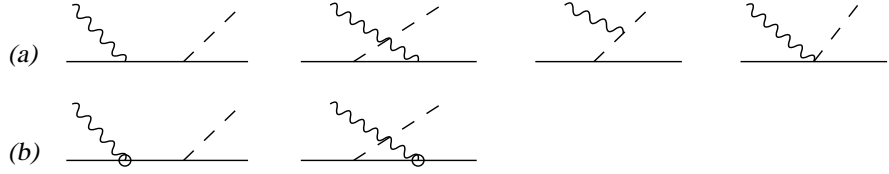}
}
\caption{Tree-level pion photoproduction graphs. The circled vertex corresponds to
the Pauli coupling.}
\figlab{Born_chpt}
\end{figure}

To lowest order in  $g$ the
photoabsorption cross section in this theory is dominated by the
single pion photoproduction graphs as displayed in \Figref{Born_chpt}.
We find for the corresponding GDH cross sections:
\begin{subequations}
\eqlab{piprod}
\bea
\De\si^{(\pi^0 p)} &=&
\frac{\pi C}{M^2 x^2}\,\left[  (2\al \bar s+1-x) \ln\frac{\al+\la}{\al-\la} -
2 \la[ x (\al-2) + \bar s (\al+2) ] \right], \qquad \\
\De\si^{(\pi^+ n)} &=& \frac{2\pi C}{M^2 x^2}\,
\left[  -\mu^2 \ln\frac{\be+\la}{\be-\la} + 2\la(\bar s\be-x\al)\right], \\
\De\si^{(\pi^0 n)} &=& 0,\\
\De\si^{(\pi^- p)} &=&
\frac{2\pi C}{M^2 x^2}\,\left[ -\mu^2 \ln\frac{\be+\la}{\be-\la} + (2\al \bar
s-1-x) \ln\frac{\al+\la}{\al-\la} -
2\bar s\la\right],
\eea
\end{subequations}
where $ C=\left(eg/4\pi\right)^2$, $\mu = m_\pi/M $, $m_\pi$ is
the pion mass, and
\begin{subequations}
\bea
&& s= M^2 + 2M\w,\,\,\, \bar s = s/M^2,\\
&& \al=(s+M^2-m_\pi^2)/2s,\\
&& \be =(s-M^2+m_\pi^2)/2s=1-\al ,\\
&& \la = (1/2s)\sqrt{s-(M+m_\pi)^2}\sqrt{s-(M-m_\pi)^2} \,.
\eea
\end{subequations}

As in the case of QED, the anomalous magnetic moment corrections
start at ${\cal O}(g^2)$, implying that the $lhs$ of the GDH
SR begins at ${\cal O}(g^4)$.  Since the tree-level cross
sections are ${\cal O}(g^2)$, we must require that \beq
\eqlab{conc1} \int\limits_{\w_{\rm th}}^\infty {d\omega\over
\omega }\, \De\sigma^{(I)} (\w) =0, \,\,\,\, \mbox{for $I=\pi^0
p,\,\pi^+ n,\, \pi^0 n, \pi^- p$}, \eeq where $\w_{\rm th} = m_\pi
(1+ m_\pi/2M)$ is the threshold of the pion photoproduction
reaction. This requirement is indeed verified for the expressions given in \Eqref{piprod} 
---the consistency of GDH SR is maintained in this theory for each of the pion production
channels.

We now turn our attention to the linearized GDH sum rule. In this
case we first introduce Pauli moments $\kappa_{0p}$ and
$\kappa_{0n}$ for the proton and the neutron, respectively. The
dependence of the cross-sections on these quantities can generally
be presented as: \bea \De\si(\w;\, \kappa_{0p},\, \kappa_{0n})
&=&\De\si(\w)
+\kappa_{0p} \,\De\si_{1p}(\w) + \kappa_{0n}\, \De\si_{1n}(\w) \nn\\
&+& \kappa_{0p}^2 \,\De\si_{2p}(\w) + \kappa_{0n}^2\,
\De\si_{2n}(\w) + \kappa_{0p}\,  \kappa_{0n}\, \De\si_{1p1n}(\w) +
\ldots. \eea Furthermore, we introduce proton and neutron
photoproduction cross sections  $\De\si^{(p)}$ and $\De\si^{(n)}$
and express the corresponding GDH SRs and their first derivatives.  
Analogous to the QED case, we obtain~:
\begin{itemize}
\item[(i)]  the GDH SRs:
\begin{subequations}
\eqlab{newsrs}
\beq
\frac{2 \pi \alpha}{M^2}\,\kappa_p^2 =\frac{1}{\pi}\int\limits_{\w_{th}}^\infty \!
\frac{d\w}{\w}\,\De\si^{(p)},\,\,\,\,
\frac{2 \pi \alpha}{M^2}\,\kappa_n^2 =\frac{1}{\pi}\int\limits_{\w_{th}}^\infty \!
\frac{d\w}{\w}\,\De\si^{(n)},
\eeq
\item[(ii)]  the {\it linearized} SRs
(valid to leading order in the coupling $g$): \beq \frac{4 \pi
\alpha}{M^2}\, \kappa_p =
\frac{1}{\pi}\int\limits_{\w_{th}}^\infty \!
\frac{d\w}{\w}\,\De\si_{1p}^{(p)} ,\,\,\,\, \frac{4 \pi
\alpha}{M^2}\, \kappa_n =
\frac{1}{\pi}\int\limits_{\w_{th}}^\infty \!
\frac{d\w}{\w}\,\De\si_{1n}^{(n)}, \eeq
 \item[(iii)] the consistency conditions (valid to leading order in the
coupling $g$): \beq \eqlab{ncons} 0 =
\frac{1}{\pi}\int\limits_{\w_{th}}^\infty \!
\frac{d\w}{\w}\,\De\si_{1n}^{(p)} ,\,\,\,\, 0  =
\frac{1}{\pi}\int\limits_{\w_{th}}^\infty \!
\frac{d\w}{\w}\,\De\si_{1p}^{(n)}. \eeq
\end{subequations}
\end{itemize}

The first derivatives of the cross-sections that enter in
\Eqref{newsrs}, to leading order in $g$, arise through the
interference of Born graphs \Figref{Born_chpt}(a) with the graphs
in \Figref{Born_chpt}(b) and we find:
\begin{subequations}
\eqlab{piprod1}
\bea
\De\si_{1p}^{(p)}\equiv \De\si_{1p}^{(\pi^0 p)}+ \De\si_{1p}^{(\pi^+ n)}&=&
\frac{\pi C}{M^2x^2}\left\{ 2x\la [4+(1 -2\al) (2+\bar s+2x)]
+2 \bar s\la (\al+2) \right.\nn\\
&-& \left. \mu^2x \ln\frac{\be+\la}{\be-\la}+(2\al \bar s+1-x)
 \ln\frac{\al+\la}{\al-\la} \right\},\\
\De\si_{1n}^{(n)}\equiv\De\si_{1n}^{(\pi^0 n)}+ \De\si_{1n}^{(\pi^- p)}&=&
\frac{\pi C}{M^2x}\left\{ 2\la (2+2x-\bar s)
+\mu^2\ln\frac{\be+\la}{\be-\la} -\ln\frac{\al+\la}{\al-\la}
\right\}, \qquad\\
\De\si_{1n}^{(p)}\equiv \De\si_{1n}^{(\pi^0 p)}+ \De\si_{1n}^{(\pi^+ n)}&=&
\frac{2\pi C}{M^2x^2}\left\{\ln\frac{\al+\la}{\al-\la} +2\la (x\be -\bar s \al)
\right\},\\
\De\si_{1p}^{(n)}\equiv\De\si_{1p}^{(\pi^0 n)}+ \De\si_{1p}^{(\pi^- p)}&=&
\frac{2\pi C}{M^2x^2}\left\{(2\bar s\al-x) \, \ln\frac{\al+\la}{\al-\la} +2\la
(x -2\bar s ) \right\}.
\eea
\end{subequations}

Using the latter two expressions we easily verify the consistency
conditions given in \Eqref{ncons}, while, employing the linearized SRs, we obtain:
\begin{subequations}
\eqlab{amms}
\bea
\kappa_p^{\rm (loop)} &=& \frac{M^2}{\pi e^2} \int\limits_{\w_{\mathrm{th}}}^{\infty}\!
\frac{d\w}{\w} \De\si_{1p}^{(p)}\nn\\
&=&\frac{g^2}{(4\pi)^2 } \left\{1 -
  \frac{\mu \,\left( 4 - 11{\mu }^2 + 3{\mu }^4 \right) }{\sqrt{1 - \quarter
{\mu }^2}}
  \arccos \frac{\mu }{2} - 6{\mu }^2+
  2{\mu }^2\left( -5 + 3\,{\mu }^2 \right) \ln \mu \right\}, \\
\kappa_n^{\rm (loop)} &=&\frac{M^2}{\pi e^2}
\int\limits_{\w_{\mathrm{th}}}^{\infty}\! \frac{d\w}{\w}
\De\si_{1n}^{( n)} =\frac{-2g^2}{(4\pi)^2 } \left\{2 -
\frac{\mu\,(2-\mu^2)}{\sqrt{1 - \quarter {\mu }^2}} \arccos
\frac{\mu }{2} - 2\mu^2 \ln \mu \right\}. \label{eq:bh}\eea
\end{subequations}

We have checked that \Eqref{amms} agrees with the
one-loop calculation done by using the standard Feynman-parameter
technique. It is interesting that, to this order,
the {\it pseudoscalar}
pion-nucleon coupling gives exactly the same result.

On the other hand,  this result does not agree with the 
covariant ChPT calculation of Ref.~\cite{KuM}, which are based 
upon the ``infrared-regularization'' procedure of Becher and Leutwyler. The
discrepancy is apparently due to the fact that the ''infrared-regularized'' loop amplitudes do not
satisfy the usual dispersion relations. Their analytic properties
in the energy plane are complicated by 
an additional cut due to explicit dependence on $\sqrt{s}$.
In other words, they do not
obey the {\em analyticity} constraint which is imposed on the
sum rule calculation.

It  is instructive to examine 
the chiral behavior of the one-loop result for the nucleon magnetic moment.
Expanding \Eqref{amms} around the chiral limit ($m_\pi=0$), which
incidentally corresponds here with the heavy-baryon expansion, we have
\bea
\eqlab{expand_amms}
\kappa_p^{\rm (loop)} &=& \frac{g^2}{(4\pi)^2 }\left\{1 -2\pi
\mu-2\,(2+5\ln\mu)\,\mu^2+\frac{21\pi}{4}\,\mu^3
+ O(\mu^4)\right\}, \\
\kappa_n^{\rm (loop)} &=&\frac{g^2}{(4\pi)^2 }
\left\{-4 +2\pi \mu-2\,(1-2\ln\mu)\,\mu^2-\frac{3\pi}{4}\,\mu^3+ O(\mu^4)
\right\}.
\eea
The term linear in pion mass (recall that $\mu=m_\pi/M$) is the well-known
leading nonanalytic (LNA) correction.
On the other hand, expanding the same expressions around the large $m_\pi$ limit
we find
\bea
\kappa_p^{\rm (loop)} &=& \frac{g^2}{(4\pi)^2 }  \,(5-4\ln\mu)\frac{1}{\mu^2} 
+ O(\mu^{-4}), \\
\kappa_n^{\rm (loop)} &=&\frac{g^2}{(4\pi)^2 } \,2 (3-4\ln\mu)\frac{1}{\mu^2} +
O(\mu^{-4}).
\eea 
What is intriguing here is that the one-loop
correction to the nucleon \amm\ for heavy quarks  behaves as
$1/m_{quark}$ (where $m_{quark} \sim m_\pi^2)$, precisely as expected from a
constituent quark-model picture.  Here this is a result of subtle cancellations
in \Eqref{amms} taking place for large values of $m_\pi$. In contrast, the 
infrared regularization procedure~\cite{KuM} gives
the result which exhibits pathological
behavior with increasing pion mass and diverges for $m_\pi =2M$.

Since the expressions in \Eqref{amms} have the
correct large $m_\pi$ behavior they should be better suited for the
chiral extrapolations of the lattice results than the usual
heavy-baryon expansions or the ``infrared-regularized'' relativistic theory. 
This point is clearly demonstrated by
\Figref{chibehavior}, where we plot the $m_\pi$-dependence of
the full [\Eqref{amms}],  heavy-baryon, and infrared-regularization~\cite{KuM}
leading order result for the magnetic moment of the proton and the neutron, 
in comparison  to recent lattice data~\cite{Zan04}.  In presenting these results 
we have added a constant shift (counter-term $\kappa_0$) to the magnetic
moment, i.e.,
\bea
\mu_p&=&(1+\kappa_{0p}+\kappa_p^{\rm (loop)})(e/2M),\\
\mu_n&=&(\kappa_{0n}+\kappa_n^{\rm (loop)}) (e/2M)
\eea
 and fitted it to the known experimental value of the magnetic
moment at the physical pion mass, $\mu_p\simeq 2.793$ and $\mu_n\simeq -1.913$, shown
by the open diamonds in the figure. For the value
of the $\pi NN$ coupling constant we have used $g^2/4\pi = 13.5$. The $m_\pi$-dependence
away off the physical point is then a prediction of the theory. The figure clearly
shows that the SR results, shown by the dotted lines,
is in a better agreement with the behavior obtained in lattice gauge
simulations.

 It is therefore tempting to use the SR results for the parametrization
of lattice data. For example, we consider the following two-parameter form:
\begin{subequations}
\eqlab{paran}
\bea
\mu_p&=&\left(1+\frac{\tilde \kappa_{0p}}{1+a_p m_\pi^2}+\kappa_p^{\rm (loop)}\right)\frac{e}{2M},\\
\mu_n&=&\left(\frac{\tilde\kappa_{0n}}{1+a_n m_\pi^2}+\kappa_n^{\rm (loop)}\right)\frac{e}{2M}\,,
\eea
\end{subequations}
where $\tilde \kappa_{0p}$ and $\tilde \kappa_{0n}$ are fixed to 
reproduce the experimental magnetic moments
at the physical $m_\pi$. The parameter $a$ can be fitted to lattice data.
The solid curves in \Figref{chibehavior} represent the result of such a single parameter
fit to the lattice data of Ref.~\cite{Zan04}
for the proton and neutron respectively, where $a_p=1.6/M^2$ and $a_n=1.05/M^2$,
$M$ is the physical nucleon mass.  

\begin{figure}[h,b,t,p]
\centerline{
  \epsfxsize=7cm
  \epsffile{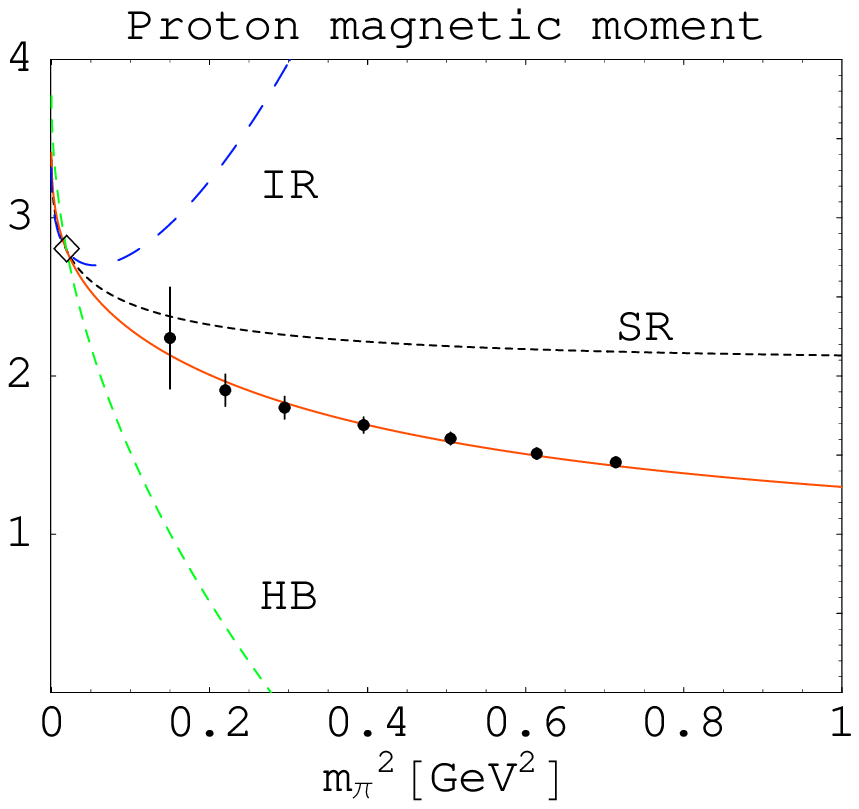}
\epsfxsize=7cm  \epsffile{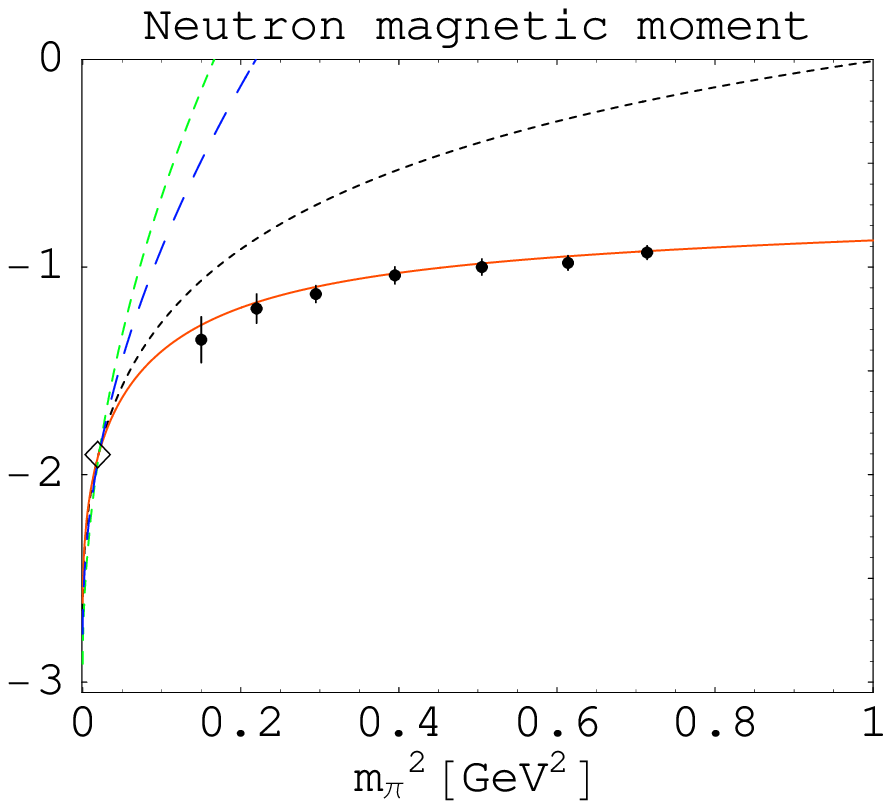}
}
\noindent
\caption{ Chiral behavior of proton and neutron magnetic moments (in nucleon magnetons) 
to one loop compared with lattice data. ``SR'' (dotted lines):
our one-loop relativistic result, ``IR'' (blue long-dashed lines): infrared-regularized relativistic
result, ``HB'' (green dashed lines): LNA term in the heavy-baryon
expansion. Red solid lines:  single-parameter  fit 
based on our SR result. Data points are results of 
lattice simulations. The open diamonds represent the experimental values
at the physical pion mass.}
\figlab{chibehavior}
\end{figure}

\section{The nucleon polarizability puzzle}
It is well-known that the Heavy-Baryon ChPT (HBChPT)
at order $p^3$ gives a remarkable prediction for the electric and magnetic polarizabilities
of the nucleon:
\begin{subequations}
\begin{eqnarray}
\eqlab{alphabet}
&&\alpha_E^{(HBLO)}=\frac{5 \pi \al }{6\, m_\pi} \left( \frac{g_A}{4
\pi f_\pi}\right)^2
=12.2 \times 10^{-4}~\mbox{fm}^3,\\ 
&&\beta_M^{(HBLO)}=\frac{\pi \al}{12\, m_\pi} 
\left( \frac{g_A}{4 \pi f_\pi}\right)^2 =1.2 \times 10^{-4}~\mbox{fm}^3\,,
\end{eqnarray}
\end{subequations}
where $g_A\simeq 1.26$, $f_\pi \simeq 93$ MeV (related to the $\pi NN$
coupling constant used in the previous section via the Goldberger-Triemann relation:
$g_A/f_\pi = g/M$). Remarkable, because this is a true prediction of HBChPT
(there are no counter-terms at this order) and because it appears to be in
a very good agreement with experiment. 

For now, I am concerned only with the sum of these polarizabilities.
On the experimental side, a recent determination
from the Baldin's sum gives~\cite{Bab98}:
\begin{eqnarray}
 \label{DDeq2.2.18}
(\alpha_E+\beta_M)_p & = &
(13.69\pm0.14)\times 10^{-4}~{\rm{fm}}^3\ , \nonumber \\
(\alpha_E+\beta_M)_n & = &
(14.40\pm0.66)\times 10^{-4}~{\rm{fm}}^3\ ,
\end{eqnarray}
for proton and neutron, respectively. 

It is also well-known that the $\De(1232)$-resonance excitation gives
a large effect to the magnetic polarizability. To quantify this effect we use
the following Lagrangian for the $\ga N\De$ coupling~\cite{PaT99,PP}:
\beq
\eqlab{ganDe}
\lag_{\ga N \De}= \frac{3\,e}{2 M (M+M_\De) }\,\ol \Psi\, T_3^\dagger
\left(i g_M  \tilde F^{\mu\nu}
- g_E \gamma_5 F^{\mu\nu}\right)\,\pa_{\mu}\psi_\nu
+ \mbox{H.c.},
\eeq
where  $\Psi$ is the nucleon field, $\psi_\mu$ is the isospin-3/2 spin-3/2 vector-spinor
field of the $\De$-isobar, $T_3$ is the isospin $N\De$ transition
matrix. The coupling constants can be deduced
from the empirical knowledge of  the $\ga N\rightarrow \De$ transition
strength. Based on the Particle Data Group  values for M1 and E2 we estimate  
$g_M \simeq 3$ and $g_E\simeq -1$.
\begin{figure}[h,t,b]
\centerline{  \epsfxsize=7 cm
  \epsffile{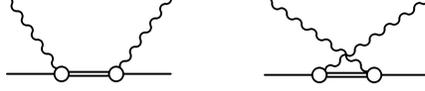}
}
\caption{The $\Delta$-excitation graphs.}
\figlab{Delta}
\end{figure} 
Computing the sum of the $s$- and $u$-channel  $\Delta$
contributions, \Figref{Delta}, to the polarizabilities one finds
(see \cite{PP} for more details):
\begin{subequations}
\bea
\al_E^{(\De)}&=&  - \frac{2\al\,g_E^2}{(M+M_\De)^3} = -0.1\times 10^{-4}~\mbox{fm}^3\\
\be_M^{(\De)} &=&  \frac{2\al g_M^2 }{(M+M_\De)^2}\frac{1}{M_\De - M}
= 7.3 \times 10^{-4}~\mbox{fm}^3.
\eea
\end{subequations}

So, while the HBChPT {\it without} Delta's is in a good agreement with experiment, HBChPT
{\it with} Delta's is not at all. Puzzling... One would expect that including
the $\De$ would improve the situation, extend the limit of applicability of our theory
higher in energy, into the resonance region.

There are suggestions voiced now and then that the effect of the $\Delta$
is canceled by the $\sigma$-meson exchange, or correlated two-pion exchange. In EFT language
this corresponds to canceling the $\Delta$ contribution by an effect which formally is
of higher order in power counting than the $\De$ contribution. This kind of scenario 
has recently been explored
in Ref.~\cite{GH04} where counter-terms of $O(p^4)$ were ``promoted'' 
to lower order in order to cancel $\be_M^{(\De)}$ . 

Here I would like to argue that possibly there is a more natural explanation within the
{\it relativistic} chiral EFT.  To find the leading order prediction of chiral loops
relativistically  
we have computed the unpolarized total cross-sections, corresponding to the
Born graphs of single-pion photoproduction~\cite{prep}:
\begin{subequations}
\bea
\si^{(\pi^0 p)} &=& 
\frac{\pi C}{M\w^3}\,\left\{ [\w^2-\mu^2 \al\, s]\,
\ln\frac{\al+\la}{\al-\la} + 2 \la \left[ \w^2 (\al-2) 1 + s \mu^2  \right] \right\} \nn\\
\si^{(\pi^+ n)} &=& \frac{2\pi C}{M\w^3}\,\left[-\be \, s\,\mu^2\ln\frac{\be+\la}{\be-\la} +2\la
\, (\al\,\w^2 +s \,\mu^2)\right], \\
\si^{(\pi^0 n)} &=& 0,\nn\\
\si^{(\pi^- p)} &=& 
\frac{2\pi C}{M\w^3}\,\left\{
\w^2 \ln\frac{\al+\la}{\al-\la}-\mu^2(s\be-\mu^2M^2) 
\,\ln\frac{\be+\la}{\be-\la}\, \frac{\al+\la}{\al-\la} + 2 s\mu^2\la \right\}.
\eea
\end{subequations}
Substituting these expressions into the Baldin SR, Eq.~(\ref{DDeq2.2.14}), we obtain:
\begin{subequations}
\bea
(\al_E+\be_M)_p^{(RLO)}&=&\frac{e^2g^2}{16\pi^3 M^3}\left\{[3(1-4\mu^2+2\mu^4)
+\mbox{$\frac{1}{3}$}\mu^2]\ln\mu+\frac{406-737\mu^2+304\mu^4-36\mu^6}{6(4-\mu^2)^2}
\right.\nn\\
&+&\left.
\frac{44-788\mu^2+1500\mu^4-899\mu^6+215\mu^8-18\mu^{10}}{3\mu(4-\mu^2)^{5/2}}\, 
\arctg{\sqrt{\frac{4}{\mu^2}-1} } \,\right\}\,,\\
(\al_E+\be_M)_n^{(RLO)}&=&\!\frac{e^2g^2}{16\pi^3 M^3}\left\{
\ln\mu+\frac{1}{(4-\mu^2)^2}\left[\frac{2(2-3\mu^2)(11-5\mu^2)-3\mu^6}{3\mu\sqrt{4-\mu^2}}\, 
\arctg{\sqrt{\frac{4}{\mu^2}-1} } +5-\mu^2\right]\right\}.\nn\\
\eea
\end{subequations}
Note that the same result is obtained in the conventional one-loop calculation~\cite{albeloop}.
The semi-relativistic (or, in this case also, chiral) expansion goes as follows:
\begin{subequations}\bea
(\al_E+\be_M)_p^{(RLO)}&=&\frac{e^2g^2}{(4\pi)^2 M^3}
\frac{11}{48\mu}\left( 1+\frac{48 (4+3\ln\mu)}{11\pi}\mu - \frac{1521}{88}\mu^2 +\ldots\right) \\
(\al_E+\be_M)_n^{(RLO)} &=&\frac{e^2g^2}{(4\pi)^2 M^3}
\frac{11}{48\mu}\left( 1+\frac{4(1+12\ln\mu)}{11\pi}\mu - \frac{117}{88}\mu^2 +\ldots\right)
\eea
or, numerically (using $g^2/4\pi=13.8$, $M=0.9383$ GeV, $\mu=0.148$),
\bea
(\al_E+\be_M)_p^{(RLO)}&=& 14.5 -5.2 - 5.5 +\ldots= 5.3 \\
(\al_E+\be_M)_n^{(RLO)}&=& 14.5 - 5.5 - 0.4 + \ldots=8.7
\eea
\end{subequations}
in the usual units. (The total values are consistent with L'vov's numerical
calculation~\cite{Lvov93},  if we use $g^2/4\pi=14.2$. From that calculation it is clear that
a lot of the reduction in the value of the sum affects the magnetic polarizability.)

Therefore, as one can see, the fully relativistic leading order result is substantially
different from the non-relativistic (heavy-baryon) limit. The relativistic corrections
which are suppressed by $m_\pi/M\simeq 1/7$, and hence are supposed to be small, appear with
large coefficients and actually are not that small. 
The good news is that this apparently allows us to accommodate the
large effect of the $\De$ isobar. In fact, the $\De$ contribution now improves the
agreement with experiment. Adding the RLO and $\De$ numbers we have:
\bea
(\al_E+\be_M)_p^{(RLO+\De)}&=& 5.3+7.2=12.5\times 10^{-4}~{\rm{fm}}^3\\
(\al_E+\be_M)_n^{(RLO+\De)}&=& 8.7+7.2=15.9\times 10^{-4}~{\rm{fm}}^3.
\eea
 At this order there is also an effect of $\pi\De$ chiral loops, but those are not yet computed relativistically.


\section{Conclusion}

The chiral EFT of QCD provides a description of the low-energy hadronic reactions
and that allows one to extract hadron properties from experiments.  On the other hand,
it predicts the chiral behavior of these properties and that allows one to make a link
to the lattice QCD calculations. These are the two fronts which at present 
make the chiral EFT indispensable in relating QCD to low-energy observables.
The purpose of this talk is to demonstrate, on simple examples of nucleon magnetic moment and
polarizabilities, that manifestly relativistic calculations do a better job than the
``heavy-baryon'' ones  on both fronts. 

The calculations done in this work were based on the real-Compton-scattering sum rules,
such as GDH and Baldin sum rules. However, the results are not different from what
one would obtained in the usual loop calculations,  provided no manipulations
(e.g., infrared regularization)
which change the analytic structure  are made. 
As is shown in the works of Gegelia {\it et al.}~\cite{Geg99}, there is no
problem with power-counting in this, straightforward, formulation of covariant ChPT, if the
renormalization scale is set in a suitable way.

\section*{Acknowledgements}
I would like to extend my gratitude to the organizers for the invitation, to all the younger
crowd for the great time, and of course to
the Sicilian mob for sparing our lives despite some tensions over our dining preferences.

This work is supported in part by DOE grant no.\
DE-FG02-04ER41302 and contract DE-AC05-84ER-40150 under
which the Southeastern Universities Research Association (SURA)
operates the Thomas Jefferson National Accelerator Facility.


\begin{thebibliography}{99}
\itemsep -2pt 

\bibitem{DPM02}
D.~Drechsel, B.~Pasquini and M.~Vanderhaeghen,
{\em Phys.\ Rept.}\  {378} (2003) 99.

\bibitem{Grab}
D.~Drechsel and L.~Tiator,
arXiv:nucl-th/0406059;
see also contributions of R.~Beck and P.~Grabmayr in these proceedings.

\bibitem{wk} G. Altarelli, N. Cabibbo, and L. Maiani, \PLB
{ 40} (1972) 415.

\bibitem{DiV01} D.A. Dicus and R. Vega, \PLB  { 501}
(2001) 44.

\bibitem{PHV04}
V.~Pascalutsa, B.~R.~Holstein and M.~Vanderhaeghen,
\PLB { 600}  (2004) 239.

\bibitem{KuM}
B.~Kubis and U.~G.~Mei\ss ner,
\NPA { 679} (2001) 698.


\bibitem{Zan04}
J.~M.~Zanotti, S.~Boinepalli, D.~B.~Leinweber, A.~G.~Williams and J.~B.~Zhang,
{\em Nucl.\ Phys.\ Proc.\ Suppl.}\  {128} (2004) 233 
[arXiv:hep-lat/0401029].


\bibitem{Bab98}
D. Babusci, G. Giordano and G. Matone,
\PRC { 57}, 291 (1998).



\bibitem{PaT99}
V.~Pascalutsa,
\PRD { 58} (1998) 096002;
V.~Pascalutsa and R.~Timmermans,
\PRC { 60} (1999) 042201.

\bibitem{PP}
V.~Pascalutsa and D.~R.~Phillips,
\PRC { 67} (2003) 055202;
\PRC { 68} (2003) 055205;
arXiv:nucl-th/0308065.


\bibitem{GH04}
R.~P.~Hildebrandt, H.~W.~Griesshammer, T.~R.~Hemmert and B.~Pasquini,
{\it Eur.\ Phys.\ J.}\ A {20} (2004) 293; see also contribution of H.~W.~Griesshammer
in these proceedings.

\bibitem{prep}
B.~R.~Holstein, V.~Pascalutsa  and M.~Vanderhaeghen, in preparation.

\bibitem{albeloop}
V.~Bernard, N.~Kaiser and  U.-G.~Mei\ss ner, \NPB\ 373 (1992)  346;
A.~Metz and D.~Drechsel, {\it Z. Phys.} A 356 (1996) 351.

\bibitem{Lvov93}
A.~I.~L'vov,
\PLB { 304}  (1993) 29.

\bibitem{Geg99}
J.~Gegelia, G.~Japaridze and X.~Q.~Wang,
{\it J.~Phys.} G {29} (2003) 2303 [arXiv:hep-ph/9910260];
T.~Fuchs, J.~Gegelia, G.~Japaridze and S.~Scherer,
\PRD { 68} (2003) 056005.

\end{thebibliography}
\end{document}